\documentclass[12pt]{article}
\usepackage{amsmath}
\usepackage{amssymb}
\begin{document}
\begin{center}
{\bf {\large{Modified Proca Theory in Arbitrary and Two Dimensions}}}

\vskip 3.4cm

{\sf  A. K. Rao$^{(a)}$, R. P. Malik$^{(a,b)}$}\\
$^{(a)}$ {\it Physics Department, Institute of Science,}\\
{\it Banaras Hindu University, Varanasi - 221 005, (U.P.), India}\\

\vskip 0.1cm

$^{(b)}$ {\it DST Centre for Interdisciplinary Mathematical Sciences,}\\
{\it Institute of Science, Banaras Hindu University, Varanasi - 221 005, India}\\
{\small {\sf {e-mails: amit.akrao@gmail.com;  rpmalik1995@gmail.com}}}
\end{center}

\vskip 3.0 cm

\noindent
{\bf Abstract:}
We demonstrate that the standard St{$\ddot u$}ckelberg-modified Proca theory (i.e. a {\it massive} Abelian 1-form theory) 
respects the {\it classical} gauge and corresponding {\it quantum} (anti-)BRST symmetry
transformations in any arbitrary dimension of spacetime within the framework of Becchi-Rouet-Stora-Tyutin 
(BRST) formalism. We further show that the St{$\ddot u$}ckelberg formalism 
gets {\it modified} in the two (1+1)-dimensions of spacetime due to a couple of discrete {\it duality} symmetries in the theory which turn out to be
responsible for the existence of the nilpotent (anti-)co-BRST symmetry transformations  corresponding to the nilpotent 
(anti-)BRST symmetry transformations of our theory. These {\it nilpotent} symmetries exist {\it together} in the modified version of the 
two (1+1)-dimensional (2D) Proca theory.
We provide the mathematical basis for the {\it modification} of the St{$\ddot u$}ckelberg technique, the
existence of the discrete {\it duality} as well as the {\it continuous} (anti-)co-BRST symmetry transformations in the 2D 
{\it modified} version of Proca theory.



\newpage

\section {Introduction}
The celebrated Proca theory (i.e., a {\it massive} Abelian 1-form theory) is the generalization of the  Maxwell (i.e., a {\it massless} Abelian 1-form)
gauge theory. Whereas the {\it latter} corresponds to a physical photon with {\it two} physical degrees of freedom, the {\it former}
describes a {\it massive} vector boson with {\it three} physical degrees of freedom in the {\it physical} four $(3+1)$-dimensions 
of spacetime. The key signature of the Maxwell theory is the existence of the first-class constraints on it (see, e.g., [1, 2]).
On the contrary, the Proca theory is endowed with the second-class constraints in the terminology of Dirac's prescription for 
the classification scheme of constraints [1, 2]. The well-known St{$\ddot {u}$ckelberg technique (see, e.g., [3] for details) converts the
second-class constraints of the Proca theory into the first-class constraints thereby restoring the beautiful gauge 
symmetry for {\it even} the {\it massive} Abelian 1-form theory where the {\it gauge invariance} 
and {\it mass} co-exist {\it together} in a beautiful fashion. 
This statement is {\it true} in any arbitrary dimension of spacetime and it is valid for {\it even}  
the higher $p$-form ($p= 2, 3...$) gauge theories. For instance, we have exploited {\it this} technique in the context of {\it massive}
Abelian 2-form gauge theory  [4, 5] where the {\it mass} and {\it gauge} symmetry co-exist {\it together}
(due to the application of the St{$\ddot {u}$ckelberg formalism).

The central purpose of our present endeavor is to show that there is a mathematically-backed precise {\it modification} 
of the St{$\ddot {u}$ckelberg technique for the $p$-form ($p= 1, 2, 3...$) {\it massive} gauge theories in $D= 2p$
dimensions of spacetime as these theories turn out to be the {\it massive} models of Hodge theory within the 
framework of Becchi-Rouet-Stora-Tyutin (BRST) formalism. To corroborate this claim, in our present endeavor,
we take up the 2D St{$\ddot {u}$ckelberg-modified Proca theory (i.e. a {\it massive} Abelian 1-form theory)
to establish that there is a precise {\it modification} of the St{$\ddot {u}$ckelberg technique [cf. Eq. (15) below]
which leads to the existence of the (anti-)co-BRST symmetries {\it corresponding} to the nilpotent (anti-) BRST 
symmetries (that exist in any arbitrary dimension of spacetime) because there is a set of discrete 
{\it duality} symmetry transformations [cf. Eqs. (23), (29)] in our 2D theory. We also show that the 
{\it physical state} of our theory is annihilated by the first-class constraints as well as their {\it dual} versions
when we choose the physical state to be the {\it harmonic} state (of the Hodge decomposed state in the quantum 
Hilbert space) that is required to be annihilated by the {\it conserved} (anti-)BRST and (anti-)co-BRST 
charges of our modified 2D Proca theory which has been proven to be an {\it interesting} model for the Hodge theory [6, 7].
We have also demonstrated that the discrete {\it duality} symmetry transformations [see, Eq. (29) below] connect the (anti-)BRST charges with the
(anti-)co-BRST charges and vice-versa. Hence, the restrictions imposed by {\it these} conserved and 
nilpotent charges on the {\it physical} state are also intimately
connected due to the {\it duality} transformations (29). These {\it restrictions} and their {\it connections} are {\it true} at the {\it tree-level} 
as well as any arbitrary loop-level diagrams in the perturbative computations for a given physical process.

The theoretical contents of our present endeavor are organized as follows. First of all, to set the notations and 
convention, we begin with the general form of the Proca theory in the next section and discuss the (anti-)BRST symmetries 
in any arbitrary dimension of spacetime for the St{$\ddot u$}ckelberg-modified version of {\it this} theory. The third section 
is devoted to the {\it modification} 
of the {\it standard} St{$\ddot {u}$ckelberg formalism in 2D for the {\it massive} Abelian 1-form (i.e. Proca) theory
where we provide the mathematical basis for the existence as well as validity of this {\it modification}. The subject matter 
of the fourth section deals with the existence of the (anti-)BRST and (anti-)co-BRST symmetries {\it together} and their
intimate relationships due to the existence of the discrete duality symmetry [cf. Eq. (29)] for the {\it modified} 
version of 2D Proca theory. Finally, in the last section, we summarize our key results  and 
point out a few future directions for further investigations in the context of higher $p$-form ($p = 2, 3...$) theories.

\section {Preliminaries: (Anti-)BRST Symmetries}

We begin with the following Lagrangian density [${\cal L}_{(P)}$] for the {\it massive} Abelian  1-form ($A^{(1)} = d x^\mu A_\mu$) vector-boson
in any arbitrary dimension of spacetime (see, e.g. [3]) 
\begin{eqnarray}
{\cal L}_{(P)} = -\frac {1}{4} F_{\mu\nu} F^{\mu\nu} +  \frac {m^2}{2}  A_\mu \, A^\mu,
\end{eqnarray}
where the field strength  tensor  $F_{\mu\nu}=\partial_\mu A_\nu - \partial_\nu A_\mu$ (with $\mu,\,\nu ... = 0, 1,2...D-1$)
has been derived from the 2-form $F^{(2)} = d A^{(1)} =  \frac{1} {2!}\, (d x^\mu \wedge  d x^\nu)\,F_{\mu\nu}$
where $d = d x^\mu\partial_\mu$ (with $d^2  = 0$) is the exterior derivative of the differential geometry [8-10] and $m$ is the 
rest mass of the vector boson. It can be readily checked that under the following {\it standard} St{$\ddot u$}ckelberg technique, the vector
bosonic field $A_{\mu}$ is {\it modified} in the following manner [3] 
\begin{eqnarray}
A_\mu \longrightarrow A_\mu \mp \frac {1}{m}\, \partial_\mu \phi,
\end{eqnarray}
where $\phi$ is a {\it pure} scalar field. Insertion of (2) into (1) leads to the following form of the Stueckelberg-modified Lagrangian 
density [${\cal L}_{(S)}$] in any arbitrary dimension of spacetime
\begin{eqnarray}
{\cal L}_{(P)} \longrightarrow {\cal L}_{(S)} = -\frac {1}{4} F_{\mu\nu} F^{\mu\nu} +  \frac {m^2}{2}  A_\mu \, A^\mu  
 \mp m \, A_\mu \partial^\mu \phi + \frac {1}{2}\partial_\mu\phi \, \partial^\mu\phi,
\end{eqnarray}
which respects [i.e. $\delta_g\, {\cal L}_{(S)} = 0$] the following continuous and infinitesimal local {\it classical} gauge 
symmetry transformations ($\delta_g$)
\begin{eqnarray}
\delta_g\, A_{\mu} = \partial_\mu\,\chi,\qquad \delta_g\,\phi = \pm\,m\,\chi,\qquad \delta_g\, F_{\mu\nu} = 0,
\end{eqnarray}
where $\chi (x)$ is the {\it local} gauge transformation parameter{\footnote{The Lagrangian density (3) 
is endowed with the first-class constraints: $\Pi^0\approx  0,\,\vec \nabla \cdot \vec E \mp m\, \Pi_\phi \approx 0 $ 
where $\Pi^0 = -\,F^{00} \approx 0$ is  the primary constraint and $\frac{\partial\,\Pi^0}{\partial\,t}=\vec \nabla  \cdot \vec E \mp m\, \Pi_\phi \approx 0$ 
is the secondary constraint with $\Pi_\phi = \dot\phi \mp m\,A_0$ as the
canonical conjugate momentum w.r.t. the $\phi$ field. These {\it first-class} constraints are present in the generator $G = \int d^{D-1}x\,
[\dot \chi\,\Pi^0 - \chi\,(\vec \nabla \cdot \vec E \mp m\,\Pi_{\phi})]$ which leads to the derivation of the {\it local} 
and infinitesimal classical gauge symmetry transformations (4) in any
arbitrary dimension of spacetime.}}. The {\it latter} can be exploited within the purview of BRST formalism to
derive the off-shell nilpotent (anti-)BRST symmetry transformations $[s_{(a)b}]$ in terms of the {\it fermionic} (anti-)ghost fields $(\bar C)C$ as (see, e.g. [7])
\begin{eqnarray}
&&s_{ab}\,A_{\mu} = \partial_\mu\,\bar C,\quad s_{ab}\,\bar C = 0, \quad s_{ab}\,C = -\,i\,B,\quad s_{ab}\,B = 0,
\quad s_{ab}\,\phi = \pm\, m\,\bar C,\nonumber\\
&&s_{b}\,A_{\mu} = \partial_\mu\, C,\qquad s_{b}\,C = 0, \quad s_{b}\,\bar C = i\,B,\qquad s_b\,B = 0,\qquad s_{b}\,\phi = \pm\, m\,C,
\end{eqnarray}
which are respected by the following (anti-)BRST invariant Lagrangian density (${\cal L}_b$) for the {\it modified} 
Proca theory in any arbitrary dimension of spacetime, namely,
\begin{eqnarray}
{\cal L}_{b} &=& {\cal L}_{(S)} + s_b\,s_{ab}\,\Big[\frac{i}{2}\, A^\mu\,A_\mu  - \frac {i}{2}\,\phi^2 - \frac {1}{2}\,\bar C\,C \Big], \nonumber\\
&\equiv &{\cal L}_{(S)} + s_b\,\Big[-\,i\,\bar C\,\{(\partial\,\cdot A) \pm m\,\phi + \frac {1}{2} \,B\}  \Big],\nonumber\\
&\equiv &{\cal L}_{(S)} + s_{ab}\,\Big[i\,C\,\{(\partial\,\cdot A) \pm m\,\phi + \frac{1}{2}\, B\}  \Big],
\end{eqnarray}
where $B$ is the Nakanishi-Lautrup auxiliary field and the fermionic 
($C^2 = 0,\, \bar C^2 = 0, \, C\,\bar C + \bar C\, C = 0$) (anti-)ghost fields $(\bar C)C$ are invoked to maintain
the {\it unitarity} in the theory. In its full blaze of glory, the Lagrangian density ${\cal L}_{b}$ looks as:  
\begin{eqnarray}
{\cal L}_{b } = -\frac {1}{4} F_{\mu\nu} F^{\mu\nu} +  \frac {m^2}{2}  A_\mu \, A^\mu  
 \mp m \, A_\mu \partial^\mu \phi + \frac {1}{2}\partial_\mu\phi \, \partial^\mu\phi \nonumber\\
+ B\,(\partial \cdot A \pm m\,\phi) + \frac {B^2}{2}
 -\,i\,\partial_\mu\,\bar C\,\partial^\mu\,C + i\,m^2\,\bar C\,C.
\end{eqnarray}
It is straightforward to check that $s_b\,{\cal L}_{b } = \partial_\mu \,[B\,\partial^\mu\,C]$ and 
$s_{ab}\,{\cal L}_{b} = \partial_\mu \,[B\,\partial^\mu\,\bar C]$ which render the  $D-$dimensional action integral 
$S = \int d^D x\,{\cal L}_{b}$ {\it invariant} under $s_{(a)b}$.

According to the celebrated Noether theorem, the above continuous, off-shell nilpotent ($s_{(a)b}^2 = 0$),
absolutely anticommuting ($s_b \;s_{ab} + s_{ab}\; s_b = 0$) and {\it infinitesimal} symmetry transformations 
lead to the following explicit expressions for the (anti-)BRST invariant ($s_{ab}\, J_{(ab)}^{\mu} = 0,\,
s_{b}\,J_{(b)}^{\mu} = 0$) Noether conserved currents ($J^{\mu}_{(r)}$ with $r = ab, b$):
\begin{eqnarray}
J^{\mu}_{(ab)} = -\,F^{\mu\nu}\,\partial_\nu\,\bar C + B\, \partial^\mu\,\bar C \pm m\,\bar C\,\partial^\mu\,\phi - m^2\,\bar C\,A^\mu,\nonumber\\
J^{\mu}_{(b)} = -\,F^{\mu\nu}\,\partial_\nu\,C + B\, \partial^\mu\,C \pm m\,C\,\partial^\mu\,\phi - m^2\,C\,A^\mu.
\end{eqnarray}
The conservation law $\partial_\mu\, J^{\mu}_{(r)} =0$  (with $r = b, ab$) can be proven by using the following Euler-Lagrange 
(EL) equations of motion (EoM), namely,
\begin{eqnarray}
&&\partial_\mu\,F^{\mu\nu} =  \partial^\nu\,B \pm m\,\partial^\nu\,\phi -\, m^2\,A^\nu,\qquad (\Box + m^2)\,C = 0,  \nonumber\\
&& \Box\,\phi \mp m\,(\partial \cdot A) = \pm m\,B,\qquad(\Box + m^2)\,\bar C = 0.  
\end{eqnarray}
The above conserved currents lead to the definition of conserved charges as
\begin{eqnarray}
Q_b = \int d^{D - 1}\,x \,J^{0}_{(b)} &=& \int d^{D - 1}\,x \, \Big[- F^{0\,i}\,(\partial_{i}\,C) 
+B\,\dot C \pm m\, C\,\dot \phi  - m^2\,A_0\,C \big]\nonumber\\ 
&\equiv& \int d^{D - 1}\,x \,\big[B\,\dot C - \dot B\,C \big],\nonumber\\
Q_{ab} = \int d^{D - 1}\,x \,J^{0}_{(ab)} &=& \int d^{D - 1}\,x \, \Big[- F^{0\,i}\,(\partial_{i}\,\bar C) 
+B\,\dot {\bar C} \pm m\, \bar C\,\dot \phi  - m^2\,A_0\,\bar C \big]\nonumber\\ 
&\equiv& \int d^{D - 1}\,x \,\big[B\,\dot {\bar C} - \dot B\,\bar C \big],
\end{eqnarray}
where we have applied the Gauss divergence theorem to drop the total {\it space} derivative terms and used: 
$\dot B = \vec \nabla \cdot \vec E + m^2\,A_0 \mp m\,\dot \phi$ [that arises from Eq. (9)] to derive the concise forms of 
conserved{\footnote {The conservation law (i.e. $\dot Q_{(a)b} = 0$) for the concise forms of the (anti-)BRST charges $Q_{(a)b} $
can be easily proven by using the EL-EoMs: $(\Box + m^2)\,B = 0,\, (\Box + m^2)\,C = 0,\, (\Box + m^2)\,\bar C = 0$ 
which emerge out from the variation of the action integral $S= \int d^{D-1}x\,{\cal L}_{b}$ defined w.r.t. the Lagrangian 
density ${\cal L}_{b}$.}} charges $Q_{(a)b}$. It is straightforward to note that the 
following {\it exact} forms of the charges w.r.t. $s_{(a)b}$ are {\it true}, namely,
\begin{eqnarray}
&& Q_b = \int d^{D - 1}\,x \,\big[  s_b\,\{ i\,\dot {\bar C}\, C - \,i\,\bar C\, \dot C\}\big] \equiv \int d^{D - 1}\,x
 \, \big[s_{ab}\,(i\, C\, \dot C)  \big],\nonumber\\
&& Q_{ab} = \int d^{D - 1}\,x \,\big[  s_{ab}\,\{ i\, {\bar C}\, \dot C - \,i\,\dot {\bar C}\, C\}\big] \equiv \int d^{D - 1}\,x
 \, \big[s_{b}\,(-\,i\, \bar C\, \dot {\bar C})  \big],
\end{eqnarray}
which establish the off-shell nilpotency and absolute anticommutativity of the conserved (anti-)BRST charges as illustrated below
\begin{eqnarray}
&&s_b\, Q_b \quad\;  = \quad  -\,i\,\{Q_b, Q_b\} = 0 \quad\; \Longrightarrow  \quad Q^{2}_{b} = 0 \quad  \,\;\Leftrightarrow \quad s^{2}_{b} = 0,\nonumber\\
&&s_{ab}\, Q_{ab} \;\; = \; -\,i\,\{Q_{ab}, Q_{ab}\} = 0 \quad\; \Longrightarrow  \quad  Q^{2}_{ab} = 0 \quad \Leftrightarrow \quad s^{2}_{ab} = 0,\nonumber\\
&&s_{ab}\, Q_b \quad  = \quad  -\,i\,\{Q_b, Q_{ab}\} = 0 \quad \Longrightarrow \quad s^{2}_{ab} = 0 \quad  \;\Leftrightarrow \quad \{Q_b, Q_{ab} \} = 0,\nonumber\\
&&s_{b}\, Q_{ab} \quad  = \quad  -\,i\,\{Q_{ab}, Q_{b}\} = 0 \quad \Longrightarrow \quad s^{2}_{b} = 0 \quad \;\; \Leftrightarrow \quad \{Q_{ab}, Q_{b} \} = 0,
\end{eqnarray}
where the basic principle behind the {\it continuous} symmetry transformations and their {\it generators} has been used. It is interesting to point out that 
the off-shell nilpotency [i.e. $Q^{2}_{(a)b} = 0$] of the conserved (anti-)BRST charges ($Q_{(a)b}$) is deeply related with the off-shell nilpotency 
[i.e. $s^{2}_{(a)b} = 0 $] of the (anti-)BRST transformations $s_{(a)b}$. However, we note that the absolute anticommutativity 
of the BRST charge {\it with} the anti-BRST
charge is connected with the nilpotency ($s^{2}_{ab} = 0$) of the anti-BRST transformations ($s_{ab}$). On the other
hand, the absolute anticommutativity of the anti-BRST charge {\it with} the BRST charge is intemately connected with the off-shell nilpotency
($s^{2}_{b} = 0$) of the BRST transformations ($s_b$). These statements are corroborated by a close and careful look at Eqs. (11) and (12).
These observations will be  very useful when we shall discuss the conserved (anti-)co-BRST charges and corresponding symmetries for the 
St$\ddot u$ckelberg-modified 2D Proca theory in the fourth section.

\section{St$\ddot u$ckelberg Formalism for the 2D Proca Theory: Mathematical Basis for Its Modification}

The 2D Proca theory is very {\it special} in the sense that the field strength tensor $F_{\mu\nu}$ has only {\it one}
existing component which is nothing but the electric field (i.e. $F_{01} = E$). The {\it latter} (i.e. the electric field E)
turns out to be a {\it pseudo-scalar} because it is a {\it single} object which changes sign under the parity symmetry transformation. Furthermore, as far as the 
mathematical aspect of the St$\ddot u$ckelberg modification [cf. Eq. (2)] in any arbitrary dimension of spacetime is concerned, we note that 
the vector field $A_\mu$ (belonging to the 1-form $A^{(1)} = d\,x^\mu\,A_\mu$) is modified by a 1-form 
$\Phi^{(1)} = d\,\Phi^{(0)} = d\,\phi \equiv d\,x^\mu\,\partial_\mu\,\phi$ where the 0-form $\Phi^{(0)} = \phi$ is a {\it pure} scalar field and 
$d = d\,x^\mu\,\partial_\mu$ is the exterior derivative of the differential geometry [8-10]. A close look at Eq. (2) shows that the $0$-form scalar
field $\phi$ has been incorporated into the St$\ddot u$ckelberg modification with a {\it mass} factor in the denominator 
on the dimensional ground (in the natural units: $\hbar= c = 1$). It is straightforward to note that, for the 2D Proca theory, the mass dimensions of $A_\mu$
and $\phi$ fields are {\it zero}. Hence, the modification in Eq. (2) is {\it correct} on the dimensional ground. 
As a passing remarks, it is interesting to point out that the field strength tensor $F_{\mu\nu}$ remains invariant under the modification (2)
due to the well-known St$\ddot u$ckelberg formalism.

As pointed out earlier, the two (1+1)-dimensional  (2D) theory is very {\it special} because we have the freedom to add/subtract a pseudo-scalar field 
$\tilde {\phi}$ in the modification (2) with some condition. In other words, an axial-vector 1-form $\tilde {\Phi}^{(1)} = 
d\,\tilde{\Phi}^{(0)} \equiv d\,x^\mu\,\partial_\mu\,\tilde{\phi}$
constructed with a pseudo-scalar field $\tilde{\phi}$ (i.e. the 0-form $\tilde{\Phi}^{(0)} = \tilde {\phi}$) is at our disposal. However, the 
1-form $\tilde{\Phi}^{(1)}$ is a pseudo-vector (i.e. axial-vector). To make it a {\it polar-vector} in 2D, 
we have to take a {\it single} Hodge duality $*$ operation on it (with the input: $\varepsilon^{\mu\nu} = -\, \varepsilon^{\nu\mu}$), as follows:
\begin{eqnarray}
 &&*\,\tilde{\Phi}^{(1)} = *\, d\,x^\mu\,\partial_\mu\,\tilde{\phi} = \varepsilon^{\mu\nu}\,d\,x_\nu\,\partial_\mu\,\tilde{\phi}\equiv d\,x^\mu
 \,(-\varepsilon_{\mu\nu} \,\partial^\nu\,\tilde {\phi}),
\end{eqnarray}
where, at this stage, the explicit expression: $-\, \varepsilon _{\mu\nu}\,\partial^\nu\,\tilde{\phi}$ 
is a {\it polar-vector} in 2D modified version of Proca theory. In the mathematical language,
the modification (2) can now be expressed, in terms of the 1-forms, as 
\begin{eqnarray}
A^{(1)} \longrightarrow A^{(1)} \mp \frac{1}{m}\, d\, \Phi^{(0)} \pm \frac{1}{m}\, *\, d\,\tilde{\Phi}^{(0)},
\end{eqnarray}
which leads to the following explicit {\it modification} of the St$\ddot u$ckelberg technique, namely,
\begin{eqnarray}
A_\mu \longrightarrow A_\mu \mp \frac {1}{m}\, \partial_\mu \phi \mp \frac{1}{m}\,\varepsilon _{\mu\nu}\,\partial^{\nu}\,\tilde{\phi}.
\end{eqnarray}
It should be pointed out that the mass term ($m$) has to be present in the {\it third} term, too, in the above 
equation on the dimensional ground because the mass
dimension of the pseudo-scalar field ($\tilde \phi$) is also {\it zero} in 2D theory. As a side remark, we note that the field-strength tenor $F_{\mu\nu}$
does {\it not} remain {\it invariant} under (15) [which was {\it invariant} under the {\it original} modification (2)].
In fact, the field-strength tensor $F_{\mu\nu}$ explicitly transforms, under the {\it modified} St{$\ddot{u}$ckelberg-technique  (15), as follows:
\begin{eqnarray}
F_{\mu\nu} \longrightarrow  F_{\mu\nu} \mp \frac{1}{m}\,\big[\varepsilon_{\nu\lambda}\, \partial_\mu\,\partial^\lambda\,\tilde\phi
-   \varepsilon_{\mu\lambda}\, \partial_\nu\,\partial^\lambda\,\tilde\phi\big].
\end{eqnarray}
The special feature of 2D theory is the observation that {\it only} $F_{01} = E$  exists for the field strength tensor $F_{\mu\nu}$. This implies that the
following is true due to (16), namely, 
\begin{eqnarray}
F_{01}\longrightarrow F_{01} \mp \,(\partial_1\,\partial_1\,\tilde\phi - \partial_0\,\partial_0\,\tilde\phi).
\end{eqnarray}
In other words, the electric field ($E = F_{01}$) of the Proca theory is modified, under the {\it modified} 2D St{$\ddot{u}$ckelberg-technique (15)
(with the choice: $\varepsilon_{01} = + 1 = \varepsilon^{10}$), as follows:
\begin{eqnarray}
E\longrightarrow  E \pm \frac{1}{m}\,\Box \,\tilde\phi, \qquad \qquad \Box = \partial_0\,\partial_0\ - \partial_1\,\partial_1.
\end{eqnarray}
Thus, ultimately, the 2D Proca Lagrangian density (${\cal L}^{(2D)}_{(P)}$) is modified to ${\cal L}^{(2D)}_{(S)}$
\begin{eqnarray}
{\cal L}^{(2D)}_{(P)} = \frac{1}{2}\, E^2 + \frac {m^2}{2}\, A_\mu\,A^\mu & \longrightarrow & {\cal L}^{(2D)}_{(S)} = \frac {1}{2}\, 
{(E \pm \frac{1}{m}\,\Box\,\tilde\phi)}^2 \pm m\,  E\,\tilde\phi - \frac {1}{2}\,\partial_\mu \,\tilde\phi\,\,\partial^\mu\,\tilde\phi \nonumber\\
&+& \frac {m^2}{2} A_\mu\, A^ \mu  \mp m \,A_\mu \,\partial^\mu\, \phi + \frac {1}{2}\, \partial_\mu\, \phi\, \partial^\mu\, \phi,
\end{eqnarray}
where ${\cal L}^{(2D)}_{(S)}$ is the 2D St{$\ddot {u}$ckelberg-modified version of the Lagrangian density for the Proca theory (modulo some {\it total} spacetime 
derivative terms). We lay emphasis on the fact that the square of the {\it first-term} of (19) leads to
\begin{eqnarray}
(E \pm \frac {1}{m}\,\Box\,\tilde{\phi})^2 = E^2 + \frac {1}{m^2}\, (\Box\,\tilde {\phi})^2 \pm \frac {2}{m}\,E\,\Box\,\tilde {\phi},
\end{eqnarray}
where the last two terms are the higher derivative terms for a 2D theory in view of the fact that 
$E= -\,\varepsilon^{\mu\nu}\,\partial_\mu\,A_\nu$ (with the inputs: $\varepsilon^{\mu\nu} = -\,
 \varepsilon^{\nu\mu},\, \varepsilon^{10} = +1 = \varepsilon_{01}$) {\it also} contains a partial derivative. 
However, there is a {\it solution} to this problem if we assume, at this stage, that
the pseudo-scalar field $\tilde\phi$ obeys the Klein-Gordon equation{\footnote { We shall see {\it later} that the 
{\it final} Lagrangian density [cf. Eqs. (22), (27)], with the replacement: $\Box\,\tilde\phi = -\,m^2\,\tilde\phi$, does lead to the derivation of the Klein-Gordon equation
[i.e.  $(\Box + m^2)\,\tilde \phi = 0$ ] for the pseudo-scalar field [cf. Eqs. (22) and (24) below] which has been used in the derivation of
(21) from (19).}}: $(\Box + m^2)\,\tilde \phi = 0$. 
Thus, ultimately, the substitution $\Box\,\tilde\phi = -\,m^2\,\tilde\phi$
leads to the following form of the Lagrangian density from the modified Lagrangian density in Eq. (19), namely,
\begin{eqnarray}
{\cal L}^{(2D)}_{(S)} = \frac {1}{2}\, {(E \mp m\,\tilde\phi)}^2 \pm m\,  E\,\tilde\phi - 
\frac {1}{2}\,\partial_\mu \,\tilde\phi\,\,\partial^\mu\,\tilde\phi 
+ \frac {m^2}{2} A_\mu\, A^ \mu  \mp m \,A_\mu \,\partial^\mu\, \phi + \frac {1}{2}\, \partial_\mu\, \phi\, \partial^\mu\, \phi,
\end{eqnarray}
which has been {\it considered} by us in our earlier works [6, 7]. However, this is for the {\it first-time}, we are able to 
derive the Lagrangian density (21) by exploiting the theoretical beauty and strength of the 2D {\it modified} St{$\ddot {u}$ckelberg
technique that has been quoted in Eq. (15).

At this crucial juncture, there are a few remarks on the theoretical beauty of the Lagrangian density (21). 
First of all, we find that the kinetic terms for the {\it pure} scalar field $(\phi)$ and pseudo-scalar field ($\tilde\phi$) 
carry {\it opposite} signs. In other words, the pseudo-scalar field ($\tilde\phi$) carries a {\it negative} kinetic 
term which is interesting from the point of view of the existence of the possible candidates for the dark matter and dark energy
(see. e.g. [11, 12]). Furthermore, such fields (which have been christened as the ``ghost" and ``phantom" fields in the realm of 
modern-day cosmology) play important roles in the cyclic, bouncing and self-accelerated cosmological models of the Universe 
(see. e.g. [13-15]). Second, the signs of {\it all} the terms of the Lagrangian density (21), along with the {\it gauge-fixing} term [3] in the 't Hooft gauge
(${\cal L}_{gf}$) [cf. Eq. (7)], namely,
\begin{eqnarray}
{\cal L}^{(2D)}_{(S)} + {\cal L}_{(gf)} &=& \frac {1}{2}\, {(E \mp m\,\tilde\phi)}^2 \pm m\,  E\,\tilde\phi - 
\frac {1}{2}\,\partial_\mu \,\tilde\phi\,\,\partial^\mu\,\tilde\phi 
+ \frac {m^2}{2} A_\mu\, A^ \mu \nonumber\\ &\mp& m \,A_\mu \,\partial^\mu\, \phi + \frac {1}{2}\, \partial_\mu\, \phi\, \partial^\mu\, \phi - \frac{1}{2}
\,(\partial\,\cdot A \pm m\,\phi)^2,
\end{eqnarray}
are {\it fixed} as it respects a set of a couple
of very useful discrete {\it duality} symmetry transformations (modulo some {\it total} spacetime derivative terms). These duality{\footnote {The 
root-cause behind (23) is the self-duality condition on 2D Abelian 1-form. In other words, we note that: 
$ *\, A^{(1)} = *\,(d\,x^{\mu}\, A_{\mu}) = d\,x^{\mu}\,(-\, \varepsilon_{\mu\nu} A^\nu) \equiv d\,x^{\mu}\,{\tilde A}_{\mu}$. The transformation
$A_\mu \to \mp\, i\, \varepsilon_{\mu\nu} A^\nu$ in (23) owes its origin to the self-duality condition where the dual-vector
${\tilde A}_{\mu} = -\, \varepsilon_{\mu\nu} A^\nu$.}} transformations
for the basic fields ($A_\mu,\,\phi,\,\tilde \phi$) along with $E $ and $(\partial \cdot A)$ are listed as follows:
\begin{eqnarray}
&&A_\mu \to \mp\, i\, \varepsilon_{\mu\nu} A^\nu, \qquad \qquad \; \phi \to \mp\, i\, \tilde \phi, 
\qquad \quad \tilde \phi \to \mp \,i \, \phi, \nonumber\\
&&E \to \pm\, i\, (\partial \cdot A), \qquad \qquad\; (\partial \cdot A) \to \pm \,i\, E.
\end{eqnarray} 
Third, we note that the Lagrangian density (22) leads to the EL-EoMs for the fields $\phi$ and $\tilde\phi$
as the Klein-Gordon equations, namely,
\begin{eqnarray}
(\Box + m^2)\,\phi = 0,\qquad\qquad (\Box + m^2)\,\tilde \phi = 0.
\end{eqnarray}
It is interesting to point out that {\it one} of the above equations [ i.e. $(\Box + m^2)\,\tilde \phi = 0$] has been used in the derivation of (22) from (19).
Fourth, we note that the {\it mass} term for the Proca field $A_\mu$ (i.e. $\frac {m^2}{2} A_\mu\, A^ \mu $) remains {\it invariant}
under the discrete {\it duality} symmetry transformations (23). As a consequence, the fields $\phi$ and $\tilde\phi$ carry the {\it same} rest mass ($m$).
Fifth, we note that the 2D {\it modified} St{$\ddot {u}$ckelberg-technique (15) remains {\it invariant} 
under the discrete {\it duality} symmetry{\footnote{Invariance of the {\it modified} St{$\ddot {u}$ckelberg-technique (15)
under the discrete symmetry transformations (23) and/or (29) demonstrates  that there is a prefect {\it duality} symmetry in the 2D modified version of 
Proca theory. In fact, we note that (15) and (23) are intertwined {\it together} in a subtle fashion.}} 
transformations (23). Finally, we observe that the gauge-fixed Lagrangian density (22) respects the local and infinitesimal {\it gauge} and 
{\it dual-gauge} symmetry transformations $\delta_{g}$ and $\delta_{dg}$, respectively, as listed below (see, e.g. [7])
\begin{eqnarray}
&&\delta_g\, A_\mu = \partial_\mu\,\Sigma, \qquad\qquad \delta_g\, \phi = \pm m\,\Sigma, 
\qquad \;\;\;\delta_g\, E = 0, \qquad\; \;\; \delta_g\, \tilde\phi = 0,\nonumber\\
&&\delta_{dg}\, A_\mu = -\,\varepsilon_{\mu\nu}\,\partial^{\nu}\Omega, \quad\quad \delta_{dg}\,
\tilde\phi = \mp m\,\Omega, \qquad \delta_{dg}\,E = \Box\,\Omega,\quad
\delta_{dg}\,\phi = 0,
\end{eqnarray}
if we impose {\it exactly} the same kinds of restrictions $(\Box + m^2)\,\Sigma = 0,\, (\Box + m^2)\,\Omega = 0$ on the {\it gauge} and {\it dual-gauge}
symmetry transformation parameters $\Sigma$ and $\Omega$ because we note that the following transformations of the Lagrangian density (22) are {\it true}, namely,
\begin{eqnarray}
\delta_{g}\, \big[{\cal L}^{(2D)}_{(S)} + {\cal L}_{(gf)}\big] &=& -\,(\partial\,\cdot A \pm m\,\phi)\,[\Box + m^2]\,\Sigma,\nonumber\\
\delta_{dg}\, \big[{\cal L}^{(2D)}_{(S)} + {\cal L}_{(gf)}\big] &=& \partial_\mu \Bigl [ m \,\varepsilon^{\mu\nu}\; 
\bigl (m \,A_\nu \,\Omega 
\pm \,\phi\, \partial_\nu \,\Omega  \bigr ) \pm m \,\tilde \phi \,\partial^\mu \,\Omega  \Bigr ] \nonumber\\
 &+&\,(E \mp m\,\tilde\phi)\,[\Box + m^2]\,\Omega.
\end{eqnarray}
We shall see in the next section that, within the framework of BRST formalism, there is no {\it outside} restrictions 
on the (anti-)ghost fields which are the {\it generalization} of the {\it classical} {\it gauge} and {\it dual-gauge}
transformation parameters $\Sigma$ and $\Omega$ to the {\it quantum} level.

\section{Comments on (Anti-)BRST and (Anti-)Co-BRST Symmetries: 2D Modified Proca Theory}

The (anti-)BRST and (anti-)co-BRST symmetry invariant Lagrangian density for the 2D {\it modified} Proca theory is the {\it generalization}
of the (anti-)BRST invariant Lagrangian density (7) (valid in any arbitrary D-dimensions of spacetime) 
as{\footnote{We would like to point out that our choice of the 2D Lagrangian density in Eq. (27) is quite different from the 
thread-bare discussions on various forms of the Lagrangian densities in our earlier work [7].}} (see, e.g. [7])
\begin{eqnarray}
{\cal L}_{(B)} &=& {\cal B}\, {(E \mp m\,\tilde\phi)} - \frac{1}{2}\,{\cal B}^2 \pm m\,  E\,\tilde\phi - 
\frac {1}{2}\,\partial_\mu \,\tilde\phi\,\,\partial^\mu\,\tilde\phi 
+ \frac {m^2}{2} A_\mu\, A^ \mu + \frac {1}{2}\, \partial_\mu\, \phi\, \partial^\mu\, \phi \nonumber\\ &\mp& m \,A_\mu \,\partial^\mu\, \phi
+ B\,(\partial\cdot A \pm m\,\phi) + \frac{1}{2}\,{B}^2  -\,i\,\partial_\mu\,\bar C\, \partial^\mu\,C + i\, m^2\,\bar C\,C,
\end{eqnarray}
where we have $(i)$ invoked Nakanishi-Lautrup auxiliary field ${\cal B}$ to linearize  the kinetic term of the Lagrangian density (22),
and $(ii)$ {\it chosen} a specific form of the Lagrangian density. In fact, there are {\it four} Lagrangian densities 
corresponding to (22) which have been discussed in our earlier work [7]. We have taken only  {\it one specific} form in (27) which is {\it different} from the 
choices taken in [7]. It is straightforward to note that the (anti-)BRST transformations (5) are {\it now} 
generalized with {\it additional} transformations 
for our 2D theory as:
\begin{eqnarray}
&&s_{ab}\,A_\mu = \partial_\mu\,\bar C,\qquad s_{ab}\,\bar C = 0, \qquad s_{ab}\,C = -\,i\,B, \qquad s_{ab}\,\phi = \pm m\,\bar C, \nonumber\\
&&s_{ab}\,B = 0,\qquad\qquad s_{ab}\,E = 0,\qquad \; s_{ab}\,{\cal B} = 0,\qquad \qquad s_{ab}\,{\tilde \phi} = 0,\nonumber\\
&&s_{b}\,A_\mu = \partial_\mu\,C,\qquad \;\; s_{b}\,C = 0,\qquad s_{b}\,\bar C = +\,i\,B, \qquad s_{b}\,\phi = \pm m\,C, \nonumber\\
&&s_{b}\,B = 0,\qquad\qquad \;\;s_{b}\,E = 0,\qquad \; s_{b}\,{\cal B} = 0,\qquad \qquad s_{b}\,{\tilde \phi} = 0.
\end{eqnarray}
Once again, we note that $s_b\,{\cal L}{_B} = \partial_\mu\,[B\,\partial^\mu\, C]$, $s_{ab}\,{\cal L}_{B} = \partial_\mu\,[B\,\partial^\mu\,\bar C]$.
Hence, the action integral $ S = \int d^{2}x\,{\cal L}_{(B)}$ remains (anti-)BRST invariant for the physical fields which 
vanish-off  as $x\rightarrow \pm \infty$ due to the application of Gauss's divergence theorem.

Before we derive the (anti-)co-BRST symmetry transformations from (28), it is interesting to point out that the 
discrete {\it duality} symmetry transformations (23) are generalized for the (anti-)BRST  and (anti-)co-BRST
{\it invariant} 2D Lagrangian density ${\cal L}_{(B)}$ as:
\begin{eqnarray}
&&A_\mu \to \mp\, i\, \varepsilon_{\mu\nu} A^\nu, \qquad \phi \to \mp\, i\, \tilde \phi, 
\qquad \tilde \phi \to \mp \,i \, \phi, \qquad B\to \mp i\, {\cal B},\quad {\cal B} \to \mp i\,B, \nonumber\\
&&C \to \pm i\,\bar C \qquad \qquad \;\; \bar C \to \pm i\, C, \quad \quad \;(\partial\,\cdot A) \to \pm i\, E \qquad E \to \pm i\, (\partial\,\cdot A).
\end{eqnarray}
In other words, we find that the Lagrangian density ${\cal L}_{(B)}$ remains invariant under (29). As a 
consequence of the discrete {\it duality} symmetry transformations (29), we find that the nilpotent (anti-)co-BRST symmetry 
transformations [$s_{(a)d}$] can be derived from the nilpotent (anti-)BRST symmetry transformations (28) as follows:
\begin{eqnarray}
&& s_{ad} \,A_\mu = - \varepsilon_{\mu\nu} \partial^\nu  C, \qquad   s_{ad}  C = 0, \qquad s_{ad} \bar C = +\, i\, 
{\cal B}, \qquad  s_{ad} {\cal B} = 0,\quad \;\;  s_{ad} \phi =  0,\nonumber\\ 
&& s_{ad} B =  0,\qquad \qquad s_{ad} \tilde \phi = \mp\, m\, C,\qquad s_{ad} (\partial \cdot A) = 0,\quad s_{ad} E = \Box \, C, \nonumber\\
&& s_{d} \,A_\mu = - \varepsilon_{\mu\nu} \partial^\nu \,\bar C, \qquad   s_{d}  \bar C = 0, \qquad s_{d} C = -\, i\, {\cal B},
 \qquad \quad s_{d} {\cal B} = 0,\qquad s_{d} \phi =  0,\nonumber\\ 
&& s_{d} B =  0,\qquad \qquad s_{d} \tilde \phi = \mp\, m\, \bar C,\qquad s_{d} (\partial \cdot A) = 0,\qquad s_{d} E = \Box \,\bar C.
\end{eqnarray}
Let us dwell a bit on the derivation of (30) from (28) {\it  and  vice-versa} by exploiting the sheer beauty of the discrete {\it duality}
symmetry transformations (29). Let us first focus on the gauge field $A_{\mu}$ which transforms under the (anti-)BRST
symmetry transformations as: $s_{ab}\, A_{\mu} = \partial_\mu\,\bar C,\, s_{b}\, A_{\mu} = \partial_\mu\,C$. 
From these nilpotent transformations, we can derive the  off-shell nilpotent (i.e. fermionic) (anti-)co-BRST symmetry transformations $ s_{ad}\, A_{\mu} = 
-\,\varepsilon_{\mu\nu}\,\partial^\nu\,C,\, s_{d}\, A_{\mu} =  -\,\varepsilon_{\mu\nu}\,\partial^\nu\,\bar C$ by using
the transformations (29) and the replacements $s_b \rightarrow s_d, \, s_{ab} \rightarrow s_{ad}$. It can be explicitly 
checked that $s_{b}\, A_{\mu} = \partial_\mu\,C$ goes to $ s_d\,(\mp i\,\varepsilon_{\mu\nu}\,A^\nu) = \partial_\mu \,(\pm i\, \bar C)$ under (29)
which, ultimately, implies that $s_{d} \,A_\mu = - \varepsilon_{\mu\nu} \partial^\nu \,\bar C$. In exactly similar fashion, we note that
$s_{ab}\, A_{\mu} = \partial_\mu\,C$ goes to $ s_{ad}\,(\mp i\,\varepsilon_{\mu\nu}\,A^\nu) = \partial_{\mu} \,(\pm i\, C)$
 under (29) and the replacement $s_{ab}\rightarrow s_{ad}$, which finally, leads to 
$s_{ad} \,A_\mu = - \varepsilon_{\mu\nu} \partial^\nu \,C$. We can repeat this exercise with 
$s_b\,\bar C = i\, B$, $s_{ab}\,C = -\, i\, B$ which, due to (29) {\it and} replacements  $s_b \rightarrow s_d, \, s_{ab} \rightarrow s_{ad}$,
lead to the derivation of $s_d\, C = -\, i\, {\cal B}$, $s_{ad}\, \bar C = +\, i\, {\cal B}$.
We lay emphasis on the fact that the reciprocal relationships {\it also} exist\footnote{{It is worthwhile to mention that, in an earlier work
 (see, for e.g. [7]), the derivation of   $s_{d} \,A_\mu = - \varepsilon_{\mu\nu} \partial^\nu \,\bar C$ from $s_{b}\, A_{\mu} = \partial_\mu\,C$
has been carried out for the {\it on-shell} nilpotent $s_d$ and $s_b$. However, the reciprocal relationship has {\it not} 
been established in [7].}}. To be precise, we can obtain $s_{b}\, A_{\mu} = \partial_\mu\,C$
from the co-BRST symmetry transformation $s_{d} \,A_\mu = - \varepsilon_{\mu\nu} \partial^\nu \,\bar C$ by the replacement 
$s_d\,\rightarrow\,s_b \,$ and the discrete {\it duality} transformations (29). In other words, we have: 
$s_b\,(\mp i\,\varepsilon_{\mu\nu}\,A^\nu) = - \varepsilon_{\mu\nu} \partial^\nu \,(\pm\,i\,C)$ from 
$s_{d} \,A_\mu = - \varepsilon_{\mu\nu} \partial^\nu \,\bar C$ which implies $s_{b}\, A_{\mu} = \partial_\mu\,C$.
Thus, we can derive (28) from (30), too. In exactly similar manner, we observe that the 
concise forms of the (anti-)co-BRST charges $Q_{(a)d}$, for our present 2D theory, can be derived from the 
concise forms of the (anti-)BRST charges (10), for $D = 2$, as 
\begin{eqnarray}
Q_d = \int d \,x \,\big[{\cal B}\,\dot{ \bar C} - \dot {\cal B}\,\bar C \big],\qquad
Q_{ad} = \int d \,x \,\big[{\cal B}\,\dot {C} - \dot {\cal B} \,C \big],
\end{eqnarray}
due to the discrete symmetry transformations (29). Hence, the nilpotent symmetry transformations (28) and (30) as well as the 
conserved charges (10) and (31) are interconnected due to the {\it duality}  symmetry transformations (29). To be more precise, in the mathematical language,
we have a set of very interesting mappings: $s_{(a)b}\leftrightarrow s_{(a)d}$ and  $Q_{(a)b}\leftrightarrow Q_{(a)d}$ due 
to the existence of a couple of very beautiful {\it duality} symmetry transformations (29) in our theory. It is an elementary exercise 
to note that we can {\it also} derive the concise forms of $Q_b$ and $Q_{ab}$ for our 2D theory from $Q_d$ and $Q_{ad}$ [cf. Eq. (31)]
by using the discrete duality symmetry transformations (29).

We end this section with a couple of  final remarks. First, the (anti-)co-BRST symmetry transformations (30), derived from the 
(anti-)BRST symmetry transformations (28), are the {\it symmetry} transformations for the Lagrangian density (27) as 
evident from the following:
\begin{eqnarray}
&&s_{ad}\,{\cal L}_{(B)} = \partial_{\mu}\, \Big[({\cal B} \pm m\, {\tilde \phi})\,\partial^\mu\,C \pm 
m \,\varepsilon^{\mu\nu}\,(\phi\,\partial_{\nu}\,C \,\mp \,m\,C\, A_{\nu}) \Big], \nonumber\\
&&s_{d}\,{\cal L}_{( B)} = \partial_{\mu}\, \Big[({\cal B} \pm m\, {\tilde \phi})\,\partial^\mu\,\bar C \pm 
m \,\varepsilon^{\mu\nu}\,(\phi\,\partial_{\nu}\,\bar C \,\pm \,m\,\bar C\, A_{\nu})\Big].
\end{eqnarray}
As a consequence of the the above observations, it is clear that the action integral $S = \int d^2 x\,{\cal L}_{(B)}$ 
remains invariant under (30) for the physical fields which vanish-off as $x\rightarrow \pm \infty$ due to Gauss's divergence theorem.
Second, we note that all the {\it exact} forms of the (anti-)BRST charges [that have been expressed in Eq. (11)] can be 
replicated in the context of conserved and nilpotent (anti-)co-BRST charges. The {\it latter} can be expressed in terms of
the nilpotent (anti-)co-BRST symmetry transformations due to the replacements $Q_b \rightarrow Q_d,\, Q_{ab} \rightarrow Q_{ad},
 \, s_b\rightarrow s_d, \,s_{ab}\rightarrow s_{ad}$ along with the {\it discrete} duality symmetry{\footnote{Though the duality 
symmetry transformations (29) are able to provide the relationships between the continuous symmetry transformations $s_{(a)b}$ 
and $s_{(a)d}$ and corresponding conserved charges, these continuous symmetries and conserved charges have their own {\it independent} 
identity as do the exterior and co-exterior derivatives of differential geometry [8-10].}} transformations
 that have been listed in Eq. (29).

\section{Conclusions}

One of the main highlights and key results of our present investigation is the {\it modification} in the St${\ddot u}$ckelberg-technique
 [cf. Eq. (15)] for our 2D {\it massive} Abelian 1-form gauge theory which has led to the derivation of the Lagrangian density (21)
that is the fulcrum of our whole discussion in our present investigation.
For our 2D {\it modified} version of Proca theory (i.e. a {\it massive} Abelian 1-form theory) the most {\it fundamental} 
symmetries are the {\it continuous} (anti-)BRST and (anti-)co-BRST transformations along  with the {\it discrete} duality
transformations (29). From these {\it fermionic} continuous  symmetries, one can define a {\it unique} bosonic continuous symmetry 
transformation. There is a {\it continuous} ghost-scale symmetry, too, in our theory. An elegant blend of {\it continuous} and
{\it discrete} symmetry transformations provide the physical realizations (see, e.g. [6, 7] for details)
 of the de Rham cohomological operators of the differential 
geometry [8-10] . The {\it discrete} duality symmetry transformations [cf. Eq. (29)], in particular,
provide the physical realization of the Hodge duality $*$ operation of the differential geometry [6, 7] 
where it has been established that $s_{(a)d} = \pm * \,s_{(a)b}\, *$. This relationship provides the analogue of the mathematical relationship
$\delta = \pm *\,d\,*$ of differential geometry that exists between the co-exterior derivative $(\delta)$ and the exterior derivative $(d)$.
In our present investigation, we have shown the direct utility of the {\it duality} transformations (29) in establishing the mapping 
$s_{(a)d} \leftrightarrow s_{(a)b}$ which has {\it not} been accomplished in [7] for the off-shell nilpotent (anti-)BRST $(s_{(a)b})$ and (anti-)co-BRST
symmetry transformations $(s_{(a)d})$. Furthermore, the Lagrangian density (27) is different from the Lagrangian density considered in [7].
The physical state of our theory is the 
{\it harmonic} state (in the Hodge decomposed state)
that is annihilated by the (anti-)BRST and (anti-)co-BRST charges which lead to the annihilation of the physical state (out of {\it total} 
states of the {\it quantum} Hilbert space) by the operator forms of the {\it first-class} constraints and their {\it dual} versions. For instance, it 
is evident that $Q_{(a)b}| \,phys>\, = 0$ implies that $B |\,phys >\, = 0$ and $\dot B |\,phys > \,= 0$. In other words, the first-class constraints
$B (= \Pi^0 \,)|\,phys > \,= 0$ and $\dot B\, (= \vec \nabla \cdot \vec E \mp m\, \Pi_\phi) \,|\,phys>\, = 0$ annihilate the physical states
which are consistent with the Dirac quantization condition{\footnote{It can be {\it also} seen that the
conditions: $B |\,phys >\, = 0$ and $\dot B |\,phys > \,= 0$ (which emerge out from the physicality criteria: $Q_{(a)b}| \,phys>\, = 0$ ) imply that:
$(\partial\,\cdot A \pm m\, \phi)\, |phys> = 0$ and $\frac{d}{d\,t}\, (\partial\,\cdot A \pm m\, \phi)\, |phys> = 0$ where 
$(\partial\,\cdot A \pm m\, \phi)$ is the {\it dual} of the quantity $(E \mp m\,\tilde\phi)$. The {\it latter} restriction
 [i.e. $(E \mp m\,\tilde\phi)\,|phys\,>\,= 0$] on the physical state emerges out from $Q_{(a)d}| \,phys>\, = 0$. The discussion on the various aspects of
constraints, gauge symmetry transformations, symmetry generators, etc., has been clarified very beautifully in a couple of earlier works (see, e.g. [16, 17]).}.}
 In exactly similar fashion, the condition: $Q_{(a)d}| \,phys> \,= 0$ leads 
to the  ${\cal B} |\,phys > \,= 0$ and $\dot {\cal B} |\,phys > \,= 0$ which imply that $(E \mp m\,\tilde\phi)\,|phys\,>\,= 0$
 and $\frac{d}{d\,t}\, (E \mp m\,\tilde\phi)\,|phys\,>\,= 0$ which are nothing but the {\it dual} versions of the first-class constraints
 [as can be seen by applying (29) on the {\it first-class} constraints]. These conditions are very sacrosanct and these are valid at any arbitrary loop-level
 diagrams in perturbation theory. In our very recent work [18], we have been able to show the importance of the annihilation of the {\it physical state} 
 by the {\it dual} versions of the first-class constraints which, ultimately, imply that the 2D ABJ anomaly term (i.e. the {\it pseudo-scalar} electric field E)
 is {\it trivial} in an \it interacting} 2D gauge theory  where there is a coupling between the gauge field and matter (Dirac) fields. 
In fact, we have proved that our 2D {\it interacting} Proca theory (with the Dirac fields) is {\it also} a tractable field-theoretic example
of Hodge theory (see, e.g. [18] for details). Our present discussion can be generalized for the {\it massive}  4D Abelian 2-form and 6D
Abelian 3-form gauge theories where there will be {\it modifications} of the celebrated St${\ddot u}$ckelberg formalism 
because the {\it above} higher $p$-form ($p = 2, 3...$) gauge theories have been proven to be the tractable examples of Hodge theory [19, 20].
It will be a nice future endeavor to discuss the  St${\ddot u}$ckelberg-modified  SUSY QED where the question of infrared problem and the 
existence of the ultralight particles as the dark matter candidates have been discussed in a very nice piece of recent work [21].

 \vskip0.5cm

\noindent
{\bf Acknowledgments}\\

\noindent
One of us (AKR) gratefully acknowledges the financial support from the {\it BHU-fellowship program} of the Banaras Hindu University (BHU),
Varanasi, under which the present investigation has been carried out. The authors express their deep sense of gratitude to the esteemed anonymous 
Reviewer for very sharp comments and useful suggestions.

 \vskip0.5cm

\noindent
{\textbf {\textit {Data availability statement:}}} No new data were created or analysed in this study.\\

\end{document}